\documentstyle[11pt]{article}

 \title{Generalizing with perceptrons in case of structured phase- and
 pattern-spaces\thanks{Based on the Diploma thesis of G.~Dirscherl,
 Regensburg 1996}}

\author{ G.Dirscherl$^1$, B.~Schottky$^2$, U.~Krey$^1$ \\ \\ $^1$
  Institut f\"ur Physik II der Universit\"at Regensburg,\\ 
  Universit\"atsstr.~31, D-93040 Regensburg, Germany\\ $^2$ Department
  of Comp.~Science and Appl.~Math.~, Aston University,\\ Birmingham B4
  7ET, UK}

  \date{revised, 4 December 1997 }

\input epsf
\newcommand{\CPa}{$\mbox{CP}_1$} \newcommand{\CPb}{$\mbox{CP}_2$}
\newcommand{\sign}{\rm sign} \newcommand{\bgl}{\begin{equation}}
  \newcommand{\egl}{\end{equation}}
\newcommand{\bga}{\begin{eqnarray}} \newcommand{\ega}{\end{eqnarray}}

\newcommand{\edn}{\frac{1}{N}}

\newcommand{\eps}{\varepsilon}

   \newcommand{\al}{\alpha}
   
   \newcommand{\epsi}{\epsilon}

  \newcommand{\beq}{\begin{equation}}
    \newcommand{\eeq}{\end{equation}}
  \newcommand{\beqa}{\begin{eqnarray}}
    \newcommand{\eeqa}{\end{eqnarray}} 
\begin{document}
\large
\maketitle

\begin{abstract} 
  We investigate the influence of different kinds of structure on the
  learning behaviour of a perceptron performing a classification task
  defined by a teacher rule. The underlying pattern distribution is
  permitted to have spatial correlations. The prior distribution for
  the teacher coupling vectors itself is assumed to be nonuniform.
  Thus classification tasks of quite different difficulty are
  included.  As learning algorithms we discuss Hebbian learning, Gibbs
  learning, and Bayesian learning with different priors, using methods
  from statistics and the replica formalism. We find that the Hebb
  rule is quite sensitive to the structure of the actual learning
  problem, failing asymptotically in most cases. Contrarily, the
  behaviour of the more sophisticated methods of Gibbs and Bayes
  learning is influenced by the spatial correlations only in an
  intermediate regime of $\alpha$, where $\alpha$ specifies the size
  of the training set. Concerning the Bayesian case we show, how
  enhanced prior knowledge improves the performance.
\end{abstract}

\section{Introduction}

In the statistical physics of neural networks one of the most important
paradigms is the {\it learning of a rule from examples},
\cite{Watkin,Opper}. The simplest case is that (i) the rule can be
represented by a ''teacher perceptron'',  while (ii) at the same time
the neural network, which tries to learn the rule, is also given by a
perceptron, called the ``student''. However, although much is known on
this generalization problem, at least for single-layer perceptrons, see
e.g.~\cite{Watkin,Opper} and references therein, two simplifying
assumptions are usually made, namely that (a) the ''rule'' itself, and
(b) the examples, are both completely random, i.e.~(a) without
correlations between the components $B_i$, $i=1,...,N$, of the teacher
perceptron's coupling vector $\vec B$ connecting the $N$ input units $i$
to the output unit, and (b) without correlations
between the components $\xi_i^\mu$ with different $i$ and/or $\mu$,
respectively, of the inputs $\vec\xi^\mu$.

In practical cases there exist of course such correlations, i.e.~both
{\it spatial} correlations (e.g.~in the `rule` $\vec B$, i.e.~between
different components $B_i$ of the teacher perceptron, and/or in the
components $\xi _i$ of the vectors $\vec \xi$ representing the inputs to
be classified by the system) and also {\it semantic} correlations
(e.g.~two different inputs $\vec\xi^\mu$ and $\vec\xi^\nu$ may represent
different 'handwritings' of the same word). Here we only mention that
storage problems with semantic correlations have been treated in
\cite{Wink1,Wink2} and concentrate in the following on {\it spatial}
correlations, by assuming that all patterns $\vec\xi^\mu$ are drawn
independently from the same non-trivial probability distribution, see
below. In context with the simpler 'storage capacity problem',  spatial
correlations have already been treated in
\cite{Wink1}-\cite{Engel}, but the 'correlated {\it
generalization} problem' itself, which is the focus of our paper,
has not yet been studied, as far as the authors know, except in a paper
of Tarkowski and Lewenstein, \cite{Tarkowski}, where only the special
case of Gibbs learning with uncorrelated teacher couplings was
discussed.

In all these papers on correlated patterns,
\cite{Wink1}-\cite{Tarkowski}, only  {\it
single-layer perceptrons} have been considered, whereas for {\it
uncorrelated patterns} the generalization problem has also been
extensively treated for {\it multilayer perceptrons}.
 Although a lot of interesting results, which may also be of practical
 relevance, have been obtained
 for these more realistic multilayer networks, see
e.g.~\cite{Schottky1,Schottky2},
  this was for uncorrelated systems and uncorrelated tasks only.
  Moreover, it has turned out in these
 and similar studies that multilayer networks cannot be treated
 successfully without a proper understanding of the behaviour of the
 {\it single-layer sub-perceptrons}, which are the building blocks of
 the multilayer systems. Therefore we concentrate here on those
 ''pre-requisite single-layer perceptrons'', treating the influence of
 spatial correlations on the generalization ability of these simplest
 neural networks. As we will see, this influence can be useful or
 detrimental, depending on the task and on the system. If possible, we
 mention explicitly in the text, or at the end in the discussion, which
 of our results can be transferred to multilayer systems and can perhaps
 be used in some kind  of 'strategy'. Nevertheless one should stress
 at this place that the single-layer perceptron itself has become
 recently a quite popular and successful classifier in so-called {\it
 support vector machines}, \cite{Vapnik}, and is more than just a toy
 model. -- Thus far the motivation of the following.

In our paper we consider exclusively the case of so-called {\it batch
  learning}, i.e.~the 'student system' is always trained with all
  examples, which are kept in mind without any preference, and is forced
  to classify not only the last training example, but {\it all} members
  of the training set correctly, whereas with the so-called ``online
  learning'' (see e.g.~\cite{Biehl}) at every training step a {\it new}
  pattern is presented to the student and the student  only uses  this
  newly added example in the training.
  Extending our work to multilayer perceptrons for 'batch learning'
   would be in fact rather expansible whereas it is much easier for
   the case of online learning. These questions are under investigation.

In the following, by analytical methods we study therefore the
generalization problem ``with spatial  structure'' as specified below;
 a ''student
perceptron'' is considered, trying to learn by batch-algorithms
a rule given
by a ''spatially structured teacher perceptron''. The set of training
examples itself is spatially structured, too, and we study, how the student
takes over the spatial correlations inherent in the training examples
 and in the teacher perceptron, and how the generalization ability
depends on these parameters as a function of the size $\al$ of the
training set.  The main problem is of course, how the spatial structure
can be used most effectively, implicitly or explicitly, by the learning
process considered. As learning algorithms we study Hebbian learning,
Gibbs learning, and Bayesian learning, using statistical methods and the
replica formalism. Although the spatial structure of the patterns and of
the teacher machine does not matter asymptotically for $\al\to\infty$ in the
two last-mentioned cases
(see below), we find that the correlations, as well as enhanced prior
information in the Bayesian case, can be quite useful at {\it
intermediate} values of $\al$.

Concerning the spatial structure considered below, we concentrate on the
basic case of {\it segmentation} -- or more general {\it
quasi-segmentation}, see below -- of the system into a finite, or
infinite, number of segments, which have a finite mutual correlation
between the activity of the neurons belonging to the same
resp.~different segments, and similarly partitioned correlations (but
with different strengths) of the synaptic couplings joining these
neurons. Real data has such correlations, and it is usually part
of preprocessing the data to detect such global dependencies, e.g.~by
{\it Principal-Component Analysis} (PCA, see e.g.~chap.~8 in
\cite{Hertz}, or \cite{Bishop}).
 Although the simplest case we consider, spatial
correlations corresponding to just two segments of equal size, is a
restriction, the basic properties can actually be investigated quite
clearly. On the other hand it is rather natural to assume similar
correlations in the classifying 'teacher rule' as well as in the
patterns; this reflects the fact that similarities in the properties of
typical patterns correspond to a similar impact on the classification
labels of the patterns. This is again a property encountered in
practice. More details are given below.

 \section{Basic Definitions} We
 consider as usual a system with binary input patterns $\vec\xi^\mu
 =(\xi_1^\mu ,...,\xi_N^\mu )$, where the $\xi_i^\mu$ are $\pm 1$.
 These input patterns generate at the teacher and student perceptrons,
 respectively, the so-called post-synaptic fields \beq
 h_B:=\frac{1}{\sqrt{N}}\sum_{i=1}^N \ B_i\xi_i=\frac{1}{\sqrt{N}}\vec
 B\cdot\vec\xi\nonumber\eeq and \beq
 h_J:=\frac{1}{\sqrt{N}}\sum_{i=1}^N \ J_i\xi_i=\frac{1}{\sqrt{N}}\vec
 J\cdot\vec\xi\,. \eeq The corresponding outputs are $\sigma_B
 :={\rm{sign}}\,h_B$, which is the ''correct output'', given by the
 teacher, and $\sigma_J :={\rm{sign}}\,h_J$. The stability of the
 student's output -- if it is correct -- is given by the positive
 quantity $\kappa :=\sigma_B {\vec{J}\cdot\vec\xi}/(|\vec J|\sqrt{N})$.

 As usual, the {\it generalization ability} $g(\al )$ is defined as the
 probability that the student, after training, produces the same output as
 the teacher on a newly added random input, which does not belong to the
 training set. Here the ''newly added random input'' is
specified as follows: It should be different from the training inputs,
but drawn from the same probability distribution,
i.e.~with the same spatial correlations (see below). The corresponding
{\it error probability} is $\epsi :=1-g(\al )$. If there are no
correlations, $\epsi$ is given as usual by the overlap $ r := (\vec
J\cdot\vec B )/({|\vec J|\cdot|\vec B|})$ of the coupling vectors of the
two perceptrons, by $g(\al )=1-(1/\pi)\arccos (r)$, see
e.g.~\cite{Watkin,Opper}. With correlations, however, the following
non-trivial pattern- and (teacher-) phase-space correlation matrices
come into play for $i,j=1,...,N$:
\beq\label{eqMatrices} C_{ij}^P\equiv
C_{ji}^P :={\langle \xi_i\xi_j\rangle}_{\vec\xi}\,\,, \quad {\rm and}
\quad C_{ij}^T\equiv C_{ji}^T :={\langle B_iB_j\rangle}_{\vec B}
\,\,.\eeq
(For $i=j$ these correlations are of course trivial,
 i.e.~$C_{ii}^P=1$, $C_{ii}^T=\vec
B^2/N$ (also =1 without restriction).)
The brackets ${\langle ... \rangle}_{\vec \xi}$ resp.~$
{\langle ... \rangle}_{\vec B}$ imply ensemble-averages with the
corresponding binomial resp.~Gaussian probability densities, e.g.
 \beq P(\vec
B)=[(2\pi)^N{\rm Det}\,{\bf C^T}]^{-1/2} \exp \left [-\frac{1}{2}
\sum_{i,j=1}^N\, B_i (C^T_{ij})^{-1}B_j\right ] \,\,.  \eeq
 In the
following we skip the sub-indexes ${\vec\xi}$ and ${\vec B}$ for
simplicity, since we additionally assume that the system is {\it
self-averaging}; i.e.~for almost all configurations of the patterns
$\vec \xi$ and of the teacher perceptron $\vec B$ considered, the same
correlation matrices $\bf C^P$ and $\bf C^T$, and also the expressions
defined below, can not only be obtained by the ensemble-averages
${\langle ...\rangle}_{\vec\xi}$ resp.~${\langle ...\rangle}_{\vec B}$,
but also for {\it fixed} realization by averaging over equivalent pairs
of sites $(i,j)$ in the limit of infinitely large systems,
$N\to\infty$, see below. Moreover, as already mentioned, we exclude
semantic correlations by requiring that for different patterns
$\vec\xi^\mu$ and $\vec\xi^\nu$ one always has $\langle \xi_i^\mu
\xi_j^\nu\rangle=0$ for $i,j=1,...,N$. 
  With these definitions  one gets additionally the important
 parameters \beq\label{eqT}
  T:=\langle (h_B)^2\rangle =N^{-1}\sum_{i,j=1}^N
 \langle B_iB_j\xi_i\xi_j \rangle=N^{-1}\sum_{i,j=1}^N
 C_{ij}^TC_{ij}^P\label{eqT1}\,\,, \eeq \beq S:= \langle (h_J)^2\rangle
 =N^{-1}\sum_{i,j=1}^N \langle J_iJ_j\xi_i\xi_j
 \rangle=N^{-1}\sum_{i,j=1}^N C_{ij}^P \langle J_iJ_j\rangle\,\,,
\label{eqS}
\eeq  and 
\beq\label{eqR} R:=\langle h_J\cdot h_B\rangle =
N^{-1}\sum_{i,j=1}^N C_{ij}^P \langle J_iB_j\rangle\,\,.
\label{eqh_J}
\eeq
Here $T$ is fixed by the ''teacher rule'' and the spatial pattern
correlations, while $S$ and $R$
 change in course of the learning process.

As already mentioned, our paper is motivated by the natural assumption
that the spatial pattern-correlations, and the phase-space correlations 
as well, i.e.~spatial correlations
in the couplings, correspond structurally to a {\it
segmented system} in a similar way as words are segmented into letters,
but recognized as a whole, \cite{Poeppel}.
Such a segmentation arises implicitly or explicitly
 in a lot of application tasks. It is
also natural to assume that pattern- and phase-space correlations are
{\it segmented in the same way}, which means that the correlation
matrices have {\it the same eigenvectors} $\vec \epsi^{\,\,k}
=(\epsi_1^k,...,\epsi_N^k)$, with $k=1,...,N$, although the
corresponding eigenvalues $C_k^P$ and $C_k^T$ may be drastically
different (\cite{Monasson,Tarkowski}). In fact, only this agreement of
the eigenvectors is what we postulate in the following, when talking of
''{\it the general quasi-segmented case}''. Moreover, we often
specialize below to ''{\it the simplest segmented case}'' by making the
natural assumption of only two segments of the same size:
\beq\label{eqSegmented1} \vec{\xi}
:=(\vec\xi^{\,\,0},\vec\xi^{\,\,1})=(\xi_1^0,...,\xi_{N/2}^0,
\xi_1^1,...,\xi_{N/2}^1)\,,\eeq with \beq\label{eqSegmented2} \langle
\xi_i^0\xi_j^1\rangle =\delta_{i,j}\,c_p\quad ,\quad \langle
\xi_i^0\xi_j^0\rangle =\langle \xi_i^1\xi_j^1\rangle
=\delta_{i,j}\,\quad , \eeq and analogously $\vec B :=(\vec
B^{\,\,0},\vec B^{\,\,1})=(B_1^0,...,B_{N/2}^0,B_1^1,...,B_{N/2}^1)$
with \beq\label{eqSegmented3} \langle B_i^0B_j^1\rangle
=\delta_{i,j}\,c_t\quad ,\quad \langle B_i^0B_j^0\rangle =\langle
B_i^1B_j^1\rangle =\delta_{i,j}\quad   \eeq for
$i,j=1,...,\, N/2$. The correlation parameters $c_p$ and $c_t$ have to
be smaller than 1 in magnitude, but otherwise they can be arbitrary real
numbers. During the training process, also the {\it student} perceptron
develops a similar segmentation with a correlation parameter $c_s$.

In the ''general quasi-segmented case'', the generalization ability
 $g(\al)$ is obtained
from the three parameters T, S and R defined in eqs.~(\ref{eqT1}),
(\ref{eqS}) and (\ref{eqR}), by $g=2\int_0^\infty {\rm
  d}h_J\int_0^\infty{\rm d}h_B\,\, P(h_J,h_B)$, with\newline
$P(h_J,h_B)=(2\pi\sqrt{ST-R^2})^{-1}\exp \left [-(Sh_B^2+Th_J^2-2R\,
  h_B\,h_J)/(2(ST-R^2))\right ]$. The result is
\beq\label{eqgRST} g=1-\frac{1}{\pi}\arccos \left (\frac{R}{\sqrt{S\cdot
    T}}\right ) \,\,. \eeq For the ''simplest segmented case'' defined
through Eqs.~(\ref{eqSegmented1}), (\ref{eqSegmented2}) and
(\ref{eqSegmented3}), this general result is specialized, by
evaluation of $S$, $T$ and $R$, to
\beq\label{eqGsegmented} g=1-\frac{1}{\pi}\arccos \left (\frac
{r+c_pc_d}{\sqrt{(1+c_pc_s)(1+c_pc_t)}} \right )\,\,, \eeq where
\beq\label{eqcd} c_d=\frac12\left\langle \frac{\vec B^0\cdot\vec
J^{\,1}}{|\vec B^0|\cdot|\vec J^{\,1}|} + \frac{\vec B^1\cdot \vec J^0}
{|\vec B^1|\cdot|\vec J^0|}\right\rangle\eeq is the cross-correlation
between the {\it different} segments of the student's and teacher's
coupling vectors.

\section{Hebbian learning} At first, we shortly
consider Hebbian learning, although this learning prescription generally
fails for $\alpha\to\infty$ in the presence of correlations, which is
not astonishing (see e.g.~\cite{Schulten}) and strongly contrasts to
Gibbs and Bayes learning (see below). However, as we will see, even in
the presence of correlations the results for Hebbian learning are
interesting, if the number $p:=\alpha N$ of training examples is small
compared to $N$, i.e.~for $\alpha\ll 1$.

Hebbian learning is defined by the {\it
one-shot prescription}
\beq\label{eqHebb} J_i=N^{-1/2}\sum_{\mu=1}^p\,{\rm sign}\left
  (\frac{\vec B\cdot\vec\xi^\mu}{\sqrt{N}}\right )\xi_i^\mu\,\,, \eeq
which leads for the ''general quasi-segmented case '' to 
 \beq\label{eqSHebb} S=\frac{\alpha}{N}\sum_{k=1}^N\left
  [(C_k^P)^2+\frac{2\alpha}{\pi T} C_k^T(C_k^P)^3\right ] \eeq and
\beq\label{eqRHebb} R=\frac{\alpha}{N}\left ( \frac{2}{\pi T}\right
)^{1/2} \sum_{k=1}^N C_k^T (C_k^P)^2\,\,, \eeq whereas $T$ is fixed.
Here $C_k^T$ and $C_k^P$ are the eigenvalues of the correlation
matrices of Eqn.~(\ref{eqMatrices}).  From these general results one can
evaluate the generalization ability simply via Eqn.~(\ref{eqgRST}).
 For the ''simplest segmented case''
defined by Eqs.~(\ref{eqSegmented1}), (\ref{eqSegmented2}) and
(\ref{eqSegmented3})
one obtains $g(\alpha)$ from Eqn.~(\ref{eqGsegmented});
the final result for the error-probability $\epsi=1-g$ is then
\beq\label{eqGHebb} \epsi (\al )=\frac{1}{\pi}\arccos\left [ \frac{\al
    (1+2c_pc_t+c_p^2)}{\sqrt{\al\frac{\pi}{2}(1+c_pc_t)^2(1+c_p^2)
      +\al^2(1+c_pc_t)(1+3c_pc_t+3c_p^2+c_p^3c_t)}} \right ]\,.\eeq
From this result for the ''simplest segmented case'' the following
general conclusions can be drawn:
\begin{itemize}
\item
For small $\al$, Hebbian learning is quite effective: The generalization
error $\epsi (\al)$ decreases rapidly with increasing $\al$ as
$\epsi (\al)=1/2-O(\sqrt{\al})$.
\item Moreover, one can see from Fig.~1 that the
decrease of the generalization error is faster, if the correlations are
''useful'' (i.e.~for $c_tc_p>0$); whether this is the case or not, does of
course not depend on the student, but only on the given training
 examples. I.e.~if the choice of the training examples is the teacher's
 task, he (or she) should try to give examples which are {\it in
 accordance}   with the spatial correlations  inherent in the 'rule',
 such that $c_pc_t>0$. On the other hand, what the
 student could do is to {\it monitor} the spatial correlations in the
 examples to get an estimate of $c_p$ already for rather small
 $\alpha$. Then by comparison of the 'monitored' values of
 $\epsi(\alpha)$ and $c_p$ with Eqn.~(\ref{eqGHebb}), he (or she) can
 estimate $c_t$ (i.e.~an important part of the rule to be discovered,
 which may be useful afterwards for Bayesian learning, see sections 5.2 and
5.3 below, where different priors are considered). 
Of course, for the 'general quasi-segmented case' this may be illusionary.
 \item
 However, in the limit $\al\to\infty$, the error of the Hebbian
  learning prescription does not converge to
 zero, but to \end{itemize} \beq\label{eqHebbLim} \epsilon_\infty
 :=\lim_{\al\to\infty} \epsilon (\al)= \frac{1}{\pi}\arccos\left [
 \frac{1+2c_pc_t+c_p^2}{\sqrt{(1+c_pc_t) (1+3c_pc_t+3c_p^2+c_p^3c_t)}}
 \right ]\,.\eeq This {\it residual generalization error} for Hebbian
 learning is due to the fact that the correct value for the student
 structure, $c_s=c_t$, is usually not achieved for $\alpha \to\infty$,
 although $\epsilon(\alpha )$, as obtained with the Hebb rule, decreases
 monotoneously with increasing $\alpha$. Already at this place we remark
 that, in contrast, for the Gibbs and Bayes algorithms
 $\epsilon(\alpha)$ always vanishes for $\alpha\to\infty$, and there the
 asymptotics of the limiting behaviour does even not at all depend on
 the correlations (see below).

For the Hebbian case, the behaviour of $\epsilon_\infty$ as
a function of $c_p$ for different values of $c_t$ is plotted in Fig.~2.
Obviously, with Hebbian learning, correlations in the patterns usually lead
to nonvanishing residual generalization error; moreover, as already
mentioned, an {\it opposite sign} in the correlations of patterns and
teacher vector, respectively, makes the learning task more difficult.
 (This observation will probably again transfer to more complicated
networks.) Nevertheless, for fixed $c_t$, whatever the sign of $c_tc_p$
is, and although for sufficiently small values of $|c_p|$ the error
increases $\propto |c_p|$ with increasing $|c_p|$,  there is according
to Fig.~2 finally a {\it decrease} down to 0 in the residual error, if
$|c_p|$ increases beyond a certain value, which depends on $c_t$. This
again is an important statement, which means that sufficiently strong
spatial correlations in the patterns will almost always be useful.

There are thus three limits where  with Hebbian learning and fixed $c_t$  a
vanishing resisual generalization error is achieved for $\alpha\to\infty$,
namely\par

 (i) for uncorrelated pattern spaces ($c_p=0$); the value of $c_t$ does not
matter at all in this case, as can be seen already from Eq.~(\ref{eqGHebb}),
since then~$\epsi(\al)=\pi^{-1}\arccos [1+(\pi/2\al)]^{-1/2}$, which
vanishes for
$\al \to \infty$ as $\epsi
=1/\sqrt{2\pi\al}$;\par

 (ii) for $c_p=\pm 1$, with $c_t\ne (-c_p)$; in this case the
pattern segments are identical up to $\pm 1$; this corresponds to an
effective reduction of $N$ to $N/2$, i.e.\ to a doubling of $\al$, but
otherwise the same result as for (i).\par

 (iii) for $c_t= \pm 1$, with $c_p\ne(-c_t)$; in this case one has
\newline $\epsi (\al )=\pi^{-1}\arccos \{1/
\sqrt{1+\pi(1+c_p^2)/[2\alpha(1\pm c_p)^2]}\,\}$, which behaves for
$\al\to\infty$ as $\sqrt{1+c_p^2}/[\sqrt{2\pi\alpha}\,(1\pm
c_p)]$.

In contrast to (ii) and (iii), if $c_t$ is not kept fixed, but if the point
$(c_p,c_t)=(-1,1)$ or $(1,-1$) is approached  with fixed slope $\partial
c_t/\partial c_p= -x$, then, according to Eqn.~(\ref{eqHebbLim}), the
residual error $\epsilon_\infty$ is a decreasing function of $x$ for
$0<x<\infty$, with
$\epsilon_\infty=1/2$  (which corresponds to zero generalization
ability) for $ x=0^+$, via $\epsilon_\infty =1/4$ for $x=1$,
to $\epsilon_\infty=0$ for $x\to\infty$. At $x\equiv 0$, where
$\epsilon_\infty$ vanishes, there  is thus a discontinuity.

Except (i), these are just pretty artificial cases, so the Hebb rule
fails, if correlated patterns are to be learned. 

For $c_t=0$, we have found that even the {\it modified}
 Hebb prescription of
\cite{Monasson}, which corresponds to the matrix transformation
 $\vec J\to {\bf K}\cdot \vec J$ with  ${\bf K}=({\bf
C}^P+\nu{\bf I})^{-1}$, where the pattern correlation matrix ${\bf C}^P$ is
given by Eqn.~(\ref{eqMatrices}), while $\bf I$ is the $N\times N$ unit
matrix and $\nu$ an optimization parameter, would yield at most a
$\sim$30\%-reduction of the generalization error $\epsilon (\alpha )$,
although for $\nu=\sqrt{1-c_p^2}$ also $c_s$ vanishes.

\section{Gibbs learning}
In case of Gibbs learning, the student perceptron is drawn at random from
the so-called {\it version space} $\cal V$, which consists exactly of all
perceptrons which classify the training examples correctly. Tarkowski and
Lewenstein, \cite{Tarkowski}, have treated storage and generalization of
spatially and semantically correlated patterns in perceptrons, but only for
the special case of Gibbs learning with {\it uncorrelated} teacher
couplings (${\bf C}^T={\bf I}$ in Eqn.~(\ref{eqMatrices})). We extend
their approach to ${\bf C}^T\ne{\bf I}$ and correct some of their
results (see below), using E.~Gardner's replica method,
\cite{Gardner,Kleinz}. With the teacher field $u_t:=N^{-1/2}\vec
B\cdot\vec \xi$ ($=h_B$ in Eqn.~(1)) and the different student fields
$u_a:= N^{-1/2}\vec J^a\cdot\vec \xi$, where $a=1,2,...,n$ enumerates
the replicas, one obtains for general quasi-segmentation with
Eqs.~(\ref{eqS}-\ref{eqR}) the following order parameters:
 \beqa T
:=\langle u_t^2\rangle &=&N^{-1}\sum_{k=1}^NC_k^P\tilde
B_k^2\label{eq19}\\ R_a :=\langle u_t u_a\rangle
&=&N^{-1}\sum_{k=1}^NC_k^P\tilde B_k\tilde J_k^a\label{eq20}\\ S_a
:=\langle u_a^2\rangle &=&N^{-1}\sum_{k=1}^NC_k^P(\tilde
J_k^a)^2\label{eq21}\\ \label{eq22} Q_{ab}:=\langle u_a u_b\rangle
&=&N^{-1}\sum_{k=1}^NC_k^P\tilde J_k^a\tilde J_k^b\,\,.  \eeqa Here the
$C_k^P$ are again the eigenvalues of the pattern correlation matrix
${\bf C}^P$, while $\tilde B_k$ and $\tilde J_k^a$ are the components of
$\vec B$ resp.~$\vec J^a$ in the corresponding basis; the fields $u_t$
and $u_a$ can be generated from normally distributed, independent
variables $w$, $v_t$ and $v_a$ by \beqa u_t &=&\sqrt{T-\frac{R^2}{Q}}
\,v_t-\frac{R}{\sqrt{Q}}\,w\\ u_a &=&\sqrt{S-Q} \,v_a-\sqrt{Q}\,w\,\,.
\eeqa The general result for the free energy, evaluated with the replica
trick assuming replica symmetry, which is exact in the present case, is
$F ={\rm Extr}\,[{F_1+\al F_2}]$, where the {\it energy term} $F_2$ is
\beq\label{eqGenF2} F_2=2 \int\,{\rm D}w \,H(x_1) \ln H(x_2),\eeq with
$x_1 :=R\,w\,\cdot (TQ-R^2)^{-1/2}$ and $x_2=w\,\cdot (Q/(S-Q))^{1/2}$,
where ${\rm D}w :=(2\pi)^{-1/2}\,{\rm d}w \exp (-w^2/2)$ and \,\,\,
$H(x) :=\,\,\,\int_x^\infty {\rm D}w$. The {\it entropy term} $F_1$ is
given by \beqa\label{eqGenF1} F_1&=& \ln
(2\pi)-N^{-1}\sum_{k=1}^N\left\{\ln [E+(F+H)C_k^P]\right.\nonumber \\&&
\left. +\frac{FC_k^P+G^2(C_k^P)^2C_k^T}{E+(F+H)C_k^P}\right\}\nonumber
\\ && +\frac{E}{2}+GR+\frac{HS+FQ}{2}\,\,.  \eeqa Here $E$, $F$, $G$ and
$H$ are additional order parameters conjugate to $|\vec J|$, $Q$, $R$
and $S$, so that in all (since $|\vec J|$ is fixed) $F$ has to be
optimized for seven order parameters.

For our ''simplest segmented systems'', see Eqs.~(\ref{eqSegmented1}),
(\ref{eqSegmented2}), and (\ref{eqSegmented3}), the general results from
Eqs.~(\ref{eq20}), (\ref{eq21}) and (\ref{eq22}), see also
(\ref{eqgRST}), (\ref{eqGsegmented}), (\ref{eqcd}), specialize to
 \beq
R_a=r^a+c_pc_d^a \quad ,\quad S_a=1+c_pc_s^a \quad ,\quad
Q_{ab}=q^{ab}+c_pq_d^{ab}\,\,,\eeq
with
 \beqa\label{eqR-parameters}
r^a&=&N^{-1}\vec B\cdot \vec J{\,^a}\,\, ,\,\, q^{ab}=N^{-1}\vec
J^{\,a}\cdot \vec J{\,^b}\,\, ,\,\, c_s^a=2N^{-1}\vec J^{\,\,0a}\cdot
\vec J^{\,\,1a} \,,\nonumber\\ c_d^a&=&N^{-1}(\vec B^{\,0}\cdot \vec
J^{\,1a}+\vec B^{\,1}\cdot \vec J^{\,0a})\,,\nonumber\\ 
\label{eq26}
q_d^{ab}&=&N^{-1}(\vec J^{\,0a}\cdot \vec J^{\,1b}+\vec J^{\,1a}\cdot
\vec J^{\,0b})\,.\eeqa
Concerning the free energy, with the
saddle-point approach and again with the replica symmetry assumption,
the {\it entropy term} specializes to
\beqa\label{eqF1Gibbs} F_1=\frac{1}{2}\{\ln
(2\pi) +\ln [(q-1-q_d+c_s)(q-1+q_d-c_s)]\}\nonumber\\ 
-\frac{1-q+q_dc_s-c_s^2}{(q-1-q_d+c_s)(q-1+c_d+c_s)}\nonumber\\ 
+\frac{(r^2+c_d^2)(q-1+c_tq_d-c_tc_s)+2r\,c_d(c_t-q\,c_t+c_s-q_d)}
{(q-1-q_d+c_s)(q-1+q_d+c_s)(1-c_t^2)}\,, \eeqa which depends only on
the five parameters $r$, $c_s$, $c_d$, $q$, and $q_d$, but not on
$c_p$, whereas the {\it energy} contribution specializes to
\beq\label{eqF2Gibbs} F_2=4 \int {\rm D}w\, H(x_1 w)\,\ln
H(x_2w)\,\,,\eeq with \beqa
x_1&=&\frac{r+c_pc_d}{\sqrt{(1+c_pc_t)(q+c_pq_d)-(r+c_pc_d)^2}}\\ 
x_2&=&\sqrt{\frac{q+c_pc_d}{1+c_sc_p-(q+c_pq_d)}}\,.\eeqa

Using the conditions $\partial{F}/\partial r =\partial F/\partial c_s
=\partial F/\partial c_d =\partial F/\partial q =\partial F/\partial
q_d = 0$ one obtains the evolution of all interesting quantities.

A major difference to the Hebb case can be seen from the asymptotic
behaviour for $\alpha\to\infty$: For unstructured teacher perceptron
($c_t=0)$, the entropy term can be simplified, since then $q=r$,
$q_d=c_d$ and $c_s=0$. So one gets asymptotically $c_d\to c_p\cdot
(1-r)$ and \beq\label{eqRto1} r\to 1 -\frac{1}{\al ^2 C^2
  (1-c_p^2)}\,\,,\eeq with $C=(2\pi)^{-1/2}\int {\rm d}x H(x)\ln
H(x)\approx -0.360324$.

Thus with Gibbs learning in the case $c_t=0$ a perfect overlap, and thus
perfect generalization, is reached for all values of $c_p$, in contrast
to Hebbian learning, where this was only the case when $c_p=0$ (except
some limiting cases, see above). But the prefactor of the
$1/\alpha^2$-behaviour of Eqn.~(\ref{eqRto1})  is proportional to
$(1-c_p^2)$, which means that  asymptotically for the overlap $r$, but
not for the generalization ability itself (see below), spatial pattern
correlations are still slightly detrimental for the Gibbs case with
$c_t=0$, but only for the just-mentioned prefactor, whereas the
''residual error'' itself now vanishes, in contrast to the Hebb case.

Let us concentrate on the generalization error now: In Fig.~3 this
quantity is plotted for several values of $|c_p|$ ($c_t=0$ fixed),
showing that the error becomes smaller with increasing $|c_p|$ for all
$\alpha$.  In other words: the more structured the pattern space the
easier it is to actually learn the classification task given by the
teacher rule. This is in contrast to the behaviour of $r$ (see above) but
intuitively reasonable, and can be understood a bit more thoroughly by
the following consideration:

If we perform a coordinate transformation in the pattern and phase
space to diagonalize the correlation matrix (of the patterns) we have
two eigenvalues $1\pm c_p$ determining the variance of the
corresponding sites.  This means that the sites with $1-|c_p|$ are
less significant than those with $1+|c_p|$. Thus, the student can
concentrate on the $N/2$ latter ones to learn the task. Since these
are only half as many as the whole set, learning can be performed
faster. In the extreme case of $|c_p|=1$ the dimension of the system
is effectively reduced to $N/2$, leading to a rescaling of $\alpha$
with the factor 2. It is clear that this reasoning can be transferred to
more general segmentations and more complex architectures.

The above considerations provide an alternative view on the
learning problem investigated here as well, i.e.~pattern sets which can be
decomposed into components of different magnitude. Data preprocessing
using principal component analysis techniques makes use of such
structures in practical applications \cite{Hertz,Bishop}. Thus, correlations
should be helpful in general.

Nevertheless, looking at the asymptotic behaviour for
$\alpha\to\infty$ of the generalization error, which for Gibbs learning
with $c_t=0$ is
\beq\label{eqAsymp} \lim_{\al\to\infty}\,\, \eps(\al
)=\frac{1}{\pi}\arccos (r+c_pc_d)=\frac{1}{\pi}\arccos
(1-\frac{1}{\al^2C^2})\approx \frac{0.625}{\al}\,\,, \eeq we have a result
which is independent of the pattern correlations at all.  So, for large
$\alpha$, structure in the patterns has no advantage in terms of the
generalization error. Actually, the fact that the improved
generalization ability due to structure in the pattern space is confined
to an intermediate $\alpha$-regime can easily be understood: to reach
perfect generalization, the sites with eigenvalue $1-|c_p|$, which are
less significant at first, become important for $\alpha \to\infty$ to
achieve the ultimate ''fine adjustment''.

In the case of a correlated teacher vector ($c_t\neq0$) things change
a bit.  Fig.~4 shows the dependence of the generalization error on
$c_p$ for several values of $c_t$ and fixed $\alpha=2$ (which is
something like an intermediate value). We see that structure in the
patterns can actually worsen the generalization ability, if the
structure is in the opposite direction than the teacher correlation,
i.e.~for $c_pc_t<0$. This resembles the behaviour of the Hebb rule,
where such type of learning problems are difficult as well, and again
the result can probably be transferred to more general situations:

Looking at the simultaneously diagonalized correlation matrices the
reason for this becomes clear: sites with the smaller variance
$1-|c_p|$, concerning the patterns, are related to teacher sites with
the larger eigenvalue $1+|c_t|$, and therefore their loss in
significance (due to a small value $1-|c_p|$) is somehow compensated
by the larger weights of the teacher vector.

Although not analytically shown we expect from numerical evidence
perfect generalization in the limit $\alpha\to\infty$ to be achieved
for $c_t\neq0$ as well, again with the law given in (\ref{eqAsymp}).
This means that  correlations in the system {\it asymptotically} neither
improve nor worsen the generalization behaviour if one uses good
enough learning rules.

Let us now break down the behaviour into the contributions from the
several order parameters. Fig.~5a,b shows the evolution of $r(\alpha)$
for different values of $c_p$ with $c_t=0$ and $c_t=0.9$, respectively.
For $c_t=0$ a higher correlation $|c_p|$ leads to a smaller overlap.
For $c_t=0.9$ the behaviour depends on the sign of $c_p$ as well. For
small $\alpha$ the overlap $r(\alpha)$ is larger for $c_pc_t>0$ than
for $c_pc_t<0$; but for larger values of $\alpha$ the relation is {\it
opposite}. To understand this ''crossing behaviour'' we have to notice
that the magnitude of the local fields, and so of the stability of the
patterns, is enhanced (reduced) for $c_pc_t>0$ ($c_pc_t<0$):
 \begin{itemize}
\item For $\alpha\ll 1$, a small stability (small on average) merely
leads to a small bias of the  version space away from the true teacher
vector (since the training patterns lie near the classification
boundary). The direction of this small bias is naturally such that the
$c_pc_t>0$ case yields higher overlap.
\item For $\alpha \gg 1$ the biasing effect of the small stability
  disappears, since the patterns cover the space somehow dense. On the
  other hand, for $c_pc_t >0$ the phase space of the solutions is now
  more confined ($q$ is smaller) because of the constraint of a higher
  stability (see the evolution of $q(\alpha)$ in Fig.~6) of the possible
  solutions. This leads to a smaller overlap $r$ for $c_pc_t<0$ in case
  of $\alpha \gg 1$.
   \end{itemize}

Fig.~7 shows the evolution of the student structure $c_s(\alpha)$ for
a teacher correlation $c_t=0.9$. It is interesting to see that
opposite correlations in the patterns (compared to the teacher) forces
the student to adopt the teacher structure rather rapidly with a
similar explanation as given above for the evolution of $r(\alpha)$.
 
The evolution of $c_d(\alpha)$ with $c_p$ (Fig.~8 for the case $c_t=0$)
is nonmonotonic, which generally occurs if $|c_p|>|c_t|$.
Asymptotically of course, the value $c_d=c_t$ is approached. So a high
correlation in the patterns (e.g.~$c_p \widetilde > 0.7$) induce strong
correlations of $c_d(\alpha )$ in an intermediate region around
$\alpha\sim 1$, which improve (worsen) the generalization ability in
this regime for $c_pc_t>0$ ($c_pc_t<0$).

Finally we should mention that the {\it independence} of
$\epsilon(\alpha\to\infty)$ on the pattern
correlations $c_p$,
 which we have shown analytically in  Eq.~(\ref{eqAsymp}) for $c_t=0$,
corrects a different result of Tarkowski and Lewenstein,
\cite{Tarkowski}.
 For $c_t\ne 0$ and $c_p\ne 0$, because of the large
number of order parameters, we did not yet succeed in calculating the
limiting behaviour analytically, although it is probably unchanged.
Again, in view of the results of \cite{Schottky1,Schottky2}, the result
should also apply to the more complicated multilayer architectures
treated in these papers, and should also be valid in the presence of
certain classes of noise.

In the following section we treat Bayesian learning with different
priors, while the results for Adatron learning, which leads to maximal
stability but not to optimal generalization, will be discussed in a
separate paper.

\section{Bayesian learning}

Bayesian methods are succesfully used for learning in neural networks,
see \cite{Kay1,Kay2} and \cite{Bishop}.  In this approach a pattern is
classified with the purpose to minimize the probability of a 'wrong
answer'. The framework requires the specification of a prior belief
about the possible networks and a  noise model defining their answer
behaviour.

More precisely, the {\it noise model} $ p(s|\vec J,\vec \xi) $ defines the
conditional probability of getting the answer 
 $s$  (correct or not) on a given pattern $\vec
\xi$ for a general classifying automaton $\vec J$ ranging over some sample
space. The probability $p(D|\vec J)$ of the data $D$ comprising the whole 
training set is typically given by simply
{\it multiplying} all probabilities for the single members of the training
set, i.e.~pairs of training-questions with 'correct answers',
 thus assuming that  these pairs are given independently of each other,
i.e.~without semantical correlations, whereas {\it spatial}
 correlations may be included.

The so-called {\it prior} $p(\vec J)$ defines the probability that the
vector $\vec J$ describes the automaton, before the evidence of any data is
taken into account, i.e.~on the basis of some prior knowledge.
Using the Bayes theorem we get  
\beq \label{bayth}
{\cal P}(\vec J|D)=\frac{p(\vec  J)\,p(D|\vec J )}
{{\cal P}(D)} \eeq as the {\it apostiori probability} of $\vec J$ after
absorbing the evidence of the training data.
 Here the so called {\it evidence of the model}
  ${\cal P}(D) :=\sum_{\vec J}\,
p(\vec J)\,p(D|\vec J)$ serves for normalization.
The ''most probable correct answer'' $s'$ on a {\it test}-{\it question}
 $\vec \xi '$ is
then given by the weighted majority vote due to ${\cal P}(\vec J|D)$
from (\ref{bayth}). Here again we assume that the same
 spatial correlations $C_{ij}^P$, see Eqn.~(\ref{eqMatrices}),
 apply  both to the training-questions and to the test-questions, while
 in both cases the ''correct answers'' are given by the same ''teacher
 automaton'' $\vec B$, which is not specified explicitly in
  Eqn.~(\ref{bayth}) and  principally can have an architecture
  different from that of the ''student automaton'' $\vec J$ (although in
  our case we assume the same architecture). Of course, we also assume
  that the ''student'' uses the same 'noise model' both for training and
  afterwards.

In practice, a good choice of the noise model and the prior (which
include the choice of the architecture used) is a crucial point for
getting good generalization behaviour. One possibility for proper
model selection is to calculate the 'evidence' of several possible
models, \cite{Kay1,Kay2}.

Methods from statistical mechanics can be used to investigate systems
in the thermodynamic limit, see \cite{OpperHaussler}, and concerning
model selection \cite{BruceSaad}.  The purpose of this section is to
compare the behaviour of Bayesian learning to Gibbs learning in the
case of structured spaces on the one hand, and to investigate the
influence of different priors on the other.  As priors we use

\par (1) a {\it uniform prior} over all normalized student coupling
vectors;
\par (2) a {\it restricted prior} permitting only those student weight
vectors, which have the correct (and in this case assumed as known)
correlation $c_s=c_t$.  (If a sufficient number of training examples is
given, the 'student' can  get knowledge of $c_t$ by monitoring the
spatial statistics $c_p$ of the questions posed by the teacher and
applying Hebbian learning for some time, i.e.~for finite $\alpha$, see
above.)

Since we are considering here a deterministic classification, the
appropriate ''noise model'' gives probability $1$ for the correct
answer (due to the coupling vector $\vec J$ and the perceptron mapping
rule) and $0$ otherwise.

One should stress that these choices contain a rather large amount of
prior knowledge about the possible teacher rules which is not in the
same way available in practical problems.

\subsection{Relation to the Gibbs case}

In our case it is pretty easy to derive the Bayes properties from the
already calculated quantities for the Gibbs case. This is possible
since one can construct a {\it perceptron} from the Gibbsian version
space {$\cal V$} which performs like the Bayesian classification, namely
the {\it Central-Point-Perceptron} (CP-perceptron).  If the $M$
members $\vec J_l$ of the version space carry identical {\it a-priori}
probabilities, the CP-perceptron is simply \beq\label{eqCP-Perceptron}
\vec J^{\,CP}=\lim_{M\to\infty} \frac{1}{K}\,\,\sum_{l=1}^M\,\vec
J_l\,\,\,.  \eeq Here $K$ is chosen, such that $|\vec
J^{\,\,CP}|=N^{1/2}$. Therefore \beq\label{eqK^2}
K^2=\lim_{M\to\infty}\,\edn\,\sum_{l,m=1}^M\,\vec J_l\cdot\vec
J_m\,\,=\lim_{M\to\infty}\,\, [M+M(M-1)\cdot q]\,\,.\eeq So one gets
for the overlap \beq r^{CP}=\frac{\vec J^{CP}\cdot \vec
  B}{N}=\lim_{M\to\infty} \frac{1}{NM\sqrt{q}}\,\,\sum_{l=1}^M\,\,\vec
J_l\cdot\vec B\,\,.\eeq Since $r
:=\lim_{M\to\infty}\,\,(NM)^{-1}\,\sum\,\vec J_l\cdot\,\vec B$ is the
overlap for the case of Gibbs learning, we have in this way the
simple relations \beq r^{CP}=\frac{r}{\sqrt{q}}\quad ,\quad
c_d^{CP}=\frac{c_d}{\sqrt{q}}\,\,.\eeq Additionally, one needs the
correlation between the two different segments of the CP student
perceptrons~: \beqa  c_s^{\,CP} &=& \frac{2}{N}\,(\vec
J^{\,\,CP})^0\cdot(\vec J^{\,\,CP})^1=
\lim_{M\to\infty}\frac{2}{NM^2q}\sum_{l,m=1}^M \,\,\vec J_l^{\,0}\cdot
\vec J_m^{\,1}\nonumber\\&=&\frac{Mc_s+M(M-1)q_d}{M^2q}\to
\frac{q_d}{q}\,\,.\eeqa The fact that the CP-perceptron reaches the
same generalization ability as the
Bayes classification follows  from \beq\label{eq41}
\sigma^{CP} ={\rm sign}
\left (\frac{1}{M\sqrt{qN}}\,\sum_{l=1}^M\,\vec J_l\cdot\vec\xi\right
) ={\rm sign} (\langle h_J\rangle )\eeq 
and \beq\label{eq42}
\sigma^{bayes}={\sign}\left [ \sum_{l=1}^M\,{\sign} \left (\frac
    {\,\vec J_l\cdot\vec\xi}{\sqrt{N}}\right )\right ] = {\sign} (\langle
{\sign} (h_J)\rangle )\,.\eeq In \cite{Watkin,Watkin93,OpperHaussler} it is
proved for the case of vanishing pattern- and teacher-correlations
($c_p=c_t=0$) that the generalization abilities obtained with the CP
perceptron, Eqn.~(\ref{eq41}), and the corresponding Bayes algorithm,
Eqn.(\ref{eq42}), respectively, agree for almost all $\vec\xi$ in the limit
$M\to\infty$, where additionally $M\ll N$ is assumed.
 Probably the agreement of the generalization abilities  is also true, if
pattern- and teacher-correlations are included.

We mention at this place that for {\it two-layer perceptrons}, in
contrast to the present case, the CP-auto\-maton does {\it not} reach
the generalization ability of the Bayes process, except for the parity
machine: The reason for this exception
is due to the 'chequered' structure of the mapping in the second layer
of the parity machine
(each flip of the output of only one hidden node changes the final
classification from (+1) to (-1) and {\it vice versa}): This leads to
the fact that for the parity machine
exploring the phase-space {\it around} the CP solution by
the Bayesian method gives just the same result as the CP-solution
itself. The interested reader will find more details in
\cite{Schottky1}.

In the following we call the CP solution '\CPa-perceptron' if the
uniform prior (1) is used, '\CPb-perceptron' if only students with
structure $c_s=c_t$ are permitted, prior (2).

\subsection{Uniform prior}

The learning curves $\epsilon(\alpha)$ for this prior are shown in
Fig.~9 for several $c_p$ and $c_t=0$. For comparison the performance
of the Gibbs algorithm is shown as well ($c_p=0$, Gibbs).

The improvement compared to Gibbs learning is significant and remains
asymptotically, i.e. one obtains for $c_t=0$ (and probably also for
$c_t\ne 0$) a behaviour again independent from $c_p$, namely 
\cite{OpperHaussler}: \beq
\label{eqErrBayes} \lim_{\al\to\infty}\,\epsilon^{bayes}(\al )\approx
\frac{0.44}{\al} \,\,.\eeq The influence of pattern correlations is
similar as in the Gibbs case.

Figs.~10a,b present results for the overlap $r(\al )$ between teacher- and
\CPa-perceptron for $c_t=0$ and $c_t=0.9$ as a function of $c_p$.
Here, one finds similar behaviour as in the preceding section, but now
somewhat more pronounced, namely (i) for $c_t=0$ the overlap decreases
with increasing $c_p$; (ii) for $c_t\ne 0$ there is a {\it crossing}
of the results near $\al \sim 2$, and (iii) different signs of $c_p$
and $c_t$ lead to higher values of $r$ for large $\alpha$; probably this
behaviour generalizes again to multilayer networks, see
\cite{Schottky1,Schottky2}.  Fig.~11
deals with the internal structure of the \CPa-perceptron, i.e.~the
internal overlap $c_s(\al )$ of it's two segments is presented, again
for $c_t=0.9$, for various values of $c_p$.  For
$\al\to\infty$, $c_s(\al )$ converges to the internal structure of the
teacher perceptron, i.e. $c_s(\al )\to c_t$. The most prominent difference
to the case of Gibbs learning is that here in the opposite limit $\al \to 0$
the \CPa-perceptron takes the value of the spatial correlation of the
{\it patterns}, i.e.~$c_s(\al \to 0)\to c_p$. This has already been
observed with the Hebb rule, see above, and also with maximal-stability
learning, \cite{Monasson}, in connection with the simpler {\it storage
problem}.

\subsection{Restricted prior}

Now let us look at the result if an enhanced prior knowledge is given,
i.e.\ the internal structure $c_t$ of the teacher. The Bayesian
inference based on this prior has the best possible generalization
performance since all available prior knowledge is used to minimize
the error probability.

In the averaging process defined by Eq.~(\ref{eqCP-Perceptron}) only
those members $\vec J_l$ of the version space are now taken into
account, which fulfill the constraint $c_s=c_t$, i.e.~which have the
same correlation between the segments as the teacher. In this case the
teacher is a typical member of the restricted version space, so we
have $q=r$ and $q_d=c_d$. Thus, the expression for the free energy
simplifies for the \CPb-perceptron with Eqs.~(\ref{eqF1Gibbs}) and
(\ref{eqF2Gibbs}) to \beqa F={\rm
  Extr}_{r,c_d}&\Big\{&\frac{1}{2}\,\left [\ln (2\pi )+\ln ((
  r-1+c_t-c_d)(r-1-c_t+c_d))\right ]\nonumber \\ 
&&+1+\frac{c_tc_d-r}{c_t^2-1} +4\al \int {\rm D}w\,H(x)\,\ln
H(x)\Big\}\,\,,\eeqa with
$x=(r+c_pc_d)^{1/2}(1+c_pc_t-r-c_pc_d)^{-1/2}$. Extremizing w.r.~to
$r$ and $c_d$, one gets the quantities describing ${\cal V}$ in this
case, and from them the behaviour of the \CPb-perceptron.

To check the performance we choose a high teacher correlation,
$c_t=0.9$. (Clearly, for smaller $c_t$ the expected advantage should
decrease, since the actual restriction of the prior by imposing
$c_s=c_t$ is reduced). The results in Fig.~12 show the performance of
the CP$_1$ and CP$_2$ perceptron as a function of $\al $ for  $c_t=0.9$.
For intermediate values of $\alpha$, we observe in fact a quite big
improvement of the CP$_2$ results with respect to the CP$_1$ case .

However it can be shown, \cite{Diplomarbeit}, that again
asymptotically for $\al\to\infty$ the results are the same as for the
uniform prior (1). This is a well-known effect in Bayesian learning:
For large sizes of the training set the evidence of the examples
dominates the influence of the prior, which becomes increasingly
irrelevant (as long as - in our case - the correct teacher rule is
included with finite probability).

\section{Conclusions}
We have studied the generalization properties of student perceptrons,
which try to learn a ''classification rule with spatial correlations'',
implemented by a teacher perceptron with built-in spatial correlations
between the components of the coupling vector. 'Batch learning' is used,
and the patterns are drawn from a spatially nonuniform distribution as
well, allowing correlations between different sites, which can be
different, however, from the above-mentioned spatial correlations of the
teacher.  We concentrated on the natural case of ''segmented
perceptrons'' and ''segmented patterns'', where the correlations were
those of corresponding sites in different segments, and where the
different correlation matrices involved in our formalism had at least
the same eigenvectors ('quasi-segmented systems').

Using the replica method \cite{Gardner,Opper} with a 
 replica symmetric ansatz, which is exact in this
case, we obtained the behaviour of Gibbs and
Bayesian learning in the thermodynamic limit. As a third learning
algorithm we investigated the Hebb rule, and found that in the presence
of correlations it is useful only for low loading and for exceptional limiting
cases of vanishing or extreme correlation: Otherwise there remains a
residual error for $\alpha\to\infty$. However, due to its simplicity,
the Hebb rule allows the easiest determination of the site-correlation
 measure $c_t$ of the ''teacher rule'' by  monitoring the pattern
 correlation $c_p$ and the generalization error for finite $\alpha$
 and comparing with Eqn.~(\ref{eqGHebb}).
  
On the contrary, for the Gibbs and Bayes cases we find that the structure of
the patterns and of the teacher machines does not matter asymptotically for
$\al\to\infty$, and perfect generalization is  achieved. Nevertheless
in an intermediate $\alpha$-regime the performance is quite sensitive to
correlations which can improve or worsen the generalization ability.

(We only mention at this
place that we have verified some results by numerical implementation
of the learning algorithms, which is difficult for Gibbs and Bayes
processes: We simply used small systems, where the
phase space was sampled by Monte Carlo methods; a more effective way
allowing for larger systems is suggested in a recent preprint of Berg
and Engel, \cite{Berg}.)

Difficult learning cases are those with {\it opposite correlations} in the
patterns and the teacher vector, respectively. For the Hebb rule the
residual error is high, for the other learning rules the
generalization error is high for intermediate $\alpha$.

These effects can be understood better by viewing the scenario as a
learning problem with different magnitudes for different components of
patterns and teacher vectors. This consideration relate things to
methods like principal component analysis. Here an interesting and
practical extension would be to investigate the influence of noise, whose
disturbing influence should depend on the relation between its size
and the corresponding magnitudes of pattern and teacher-vector sites,
see \cite{Schottky2} for multilayer networks with noise, but still for
uncorrelated patterns.

For the Bayesian case we investigated the influence of different
priors, showing that improved prior knowledge (e.g.~based on a knowledge
of the just mentioned quantity $c_t$) enhances the performance, but
again only for an intermediate regime of $\alpha$. This corresponds to
the well known fact that prior information looses significance for large
training sets.

The case of Maximum-Stability learning, where the AdaTron algorithm of
Anlauf and Biehl provides a fast and effective learning algorithm,
\cite{Anlauf}, and a related cavity method, will be the themes of a
following paper.

\subsection*{Acknowledgements}
The authors would like to thank F. Gerl, M. Probst, J. Winkel, H.
Kirschner and B. Vogl for useful discussions.

\newpage 

\subsection*{Figure Captions} 

Fig.~1: For Hebbian learning with a  correlation parameter $c_t=0.7$ of the
two segments of the teacher perceptron, the generalization error
$\epsilon(\alpha )$ is presented as a function of the reduced size 
$\alpha :=p/N$ of the training set
 for different values of the pattern correlation parameter $c_p$.

Fig.~2: The limit of the generalization error for $\alpha\to\infty$ in case
of Hebbian learning is presented as a function of the pattern correlation
parameter $c_p$ for different values of the  correlation $c_t$ of the two
segments of the teacher perceptron.

Fig.~3: For the case of Gibbs learning, the generalization error
$\epsilon(\alpha )$ is presented as a function of the reduced size $\alpha
:=p/N$  of the training set, for $c_t=0$ and
different values of $|c_p|$.

Fig.~4: For the case of Gibbs learning and $\alpha=2$, the generalization
error $\epsilon(c_p)$ is presented as a function of the pattern correlation
parameter $c_p$ for different values of $c_t$.

Fig.~5a,b: For Gibbs learning with $c_t=0$ and $c_t=0.9$, respectively, the
normalized overlap $r(\alpha )$ of the coupling vectors of the teacher's and
the student's perceptron is presented as a function of the reduced size
$\alpha :=p/N$ of the training set.

Fig.~6: For Gibbs learning with $c_t=0.9$, the order parameters $q(\alpha
)$, which is the typical overlap between the coupling vectors of two
different student perceptrons, and $r(\alpha )$, which is the overlap
between the coupling vectors of a typical student and the teacher, are
presented as a function of the reduced size $\alpha :=p/N$ of the training
set for the two cases of $c_p=\pm 0.9$.

Fig.~7: For Gibbs learning, the evolution of the correlation parameter
$c_s(\alpha )$ between the two segments of the student perceptron, as it
develops as a function of the reduced size $\alpha :=p/N$ of the training
set, is presented over
$\alpha$ for $c_t=0.9$ and $c_p=0,\,\pm 0.7$ and $\pm 0.9$.

Fig.~8: For Gibbs learning, the evolution of the  cross-correlation
parameter $c_d$ of two different segments of the teacher's and the student
perceptron, see Eqn. (13), as it develops as a function of the reduced size
$\alpha :=p/N$ of the training set, is presented over $\alpha$ for $c_t=0.9$
and $c_p=0.2,\, 0.7$ and $0.9$.

Fig.~9: For Bayesian learning with {\it uniform prior}, i.e.~the
CP$_1$ perceptron, and $c_t=0$, the generalization error $\epsilon (\alpha
)$ is presented over the reduced size $\alpha :=p/N$ of the training set,
for $c_p=0$, $0.7$ and
$0.9$, and for comparison also for Gibbs learning with $c_p=0$.

Fig.~10a,b: For Bayesian learning with {\it uniform prior}, i.e.~the CP$_1$
perceptron, for the two cases  $c_t=0$ and $c_t=0.9$, the overlap $r(\alpha
)$ between the coupling vector of the teacher and the CP$_1$ student
perceptron is presented as a function of the reduced size $\alpha :=p/N$ of
the training set, for pattern-correlations $c_p=0,\,\pm 0.7$ and $\pm 0.9$.

Fig.~11: For Bayesian learning with {\it uniform prior}, i.e.~the CP$_1$
perceptron, for  $c_t=0.9$, the evolution of the correlation $c_s(\alpha
)$ between the two different segments of the CP$_1$ student perceptron
 is presented, as it evolves  as a function of the reduced size $\alpha
:=p/N$ of the training set, for pattern-correlations $c_p=0,\,0.2,\,0.7$ and
$0.9$.

Fig.~12: For Bayesian learning with {\it restricted prior}, i.e.~the CP$_2$
perceptron, for  $c_t=0.9$, the generalization error $\epsilon (\alpha )$
  is presented as a function of the reduced size $\alpha :=p/N$ of the
training set, for pattern correlations $c_p=0,\,0.7,\,0.9$, and also, for
comparison, with unrestricted prior (i.e.~the CP$_1$ perceptron) and
$c_p=0$.
{\topmargin0mm \textheight22cm \newpage
\epsfxsize=15cm
\epsfbox{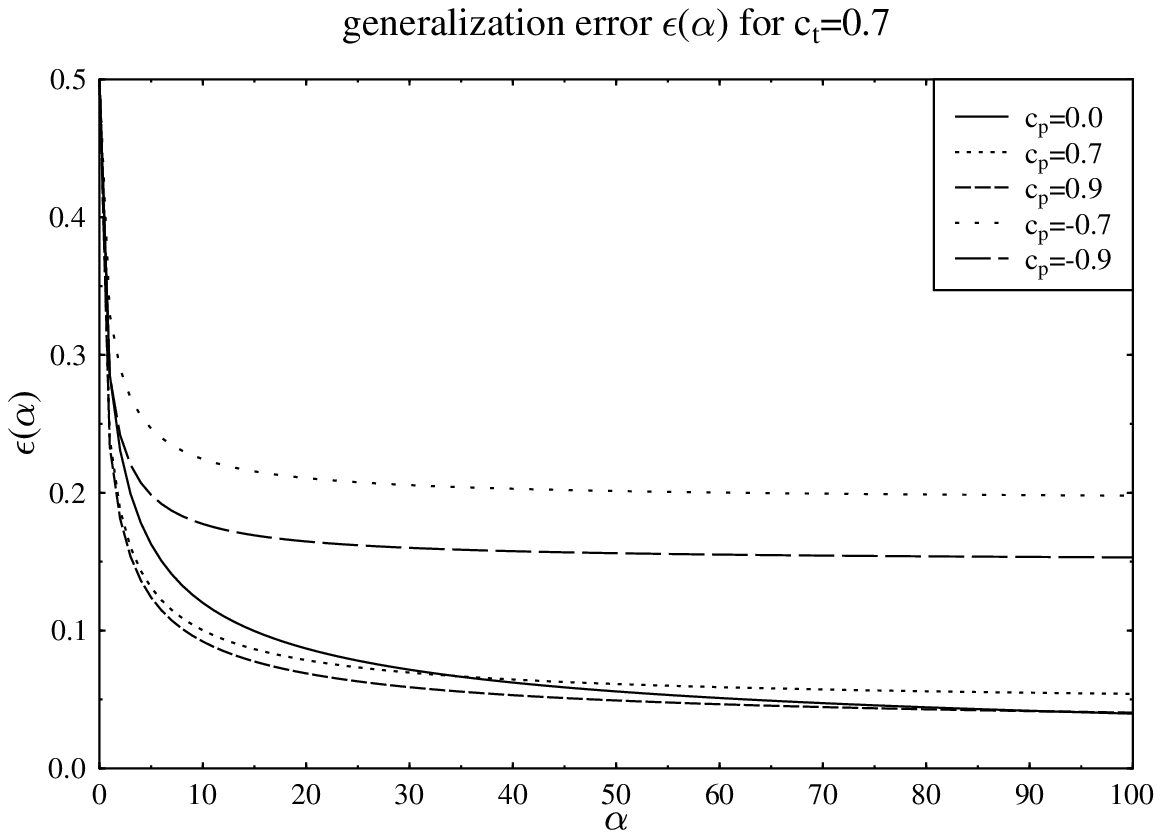}
\centerline{\underbar{Fig.~1}} 
\epsfxsize=15cm
\epsfbox{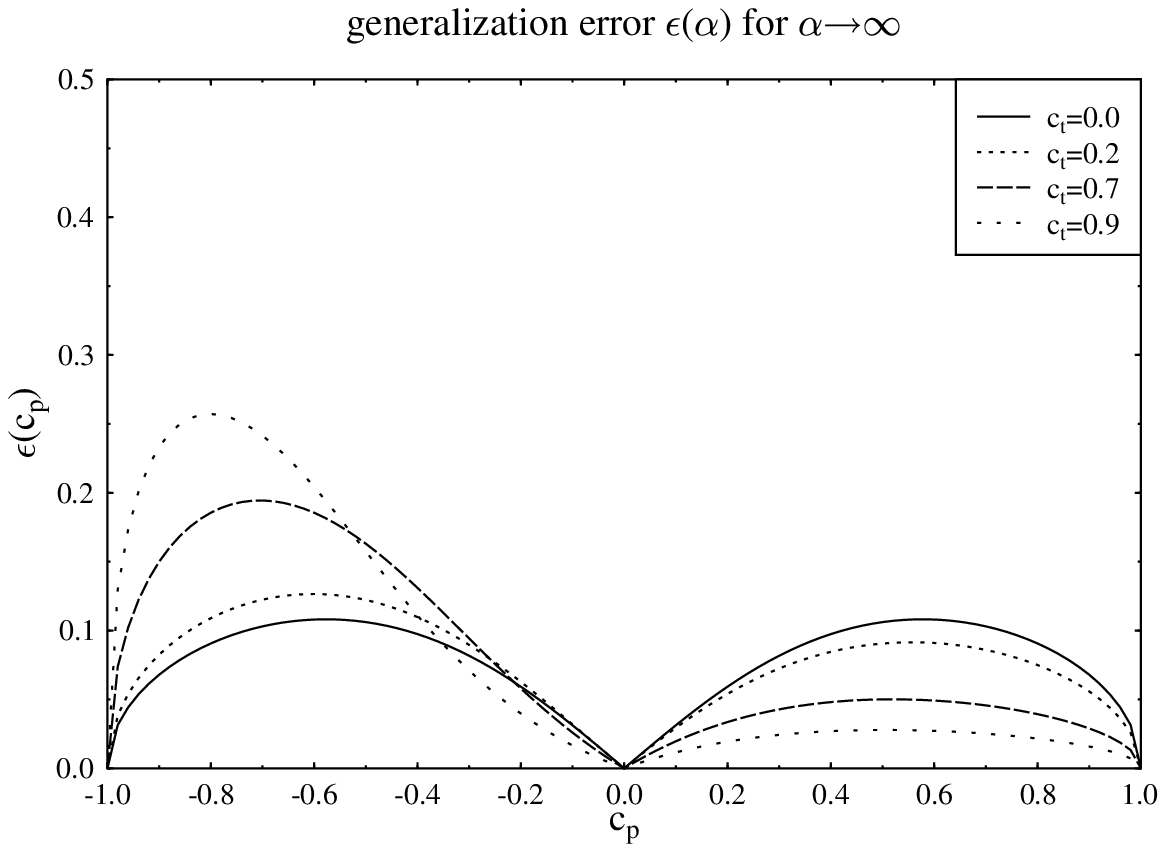}
 \centerline{\underbar{Fig.~2}}
 \epsfxsize=15cm
\epsfbox{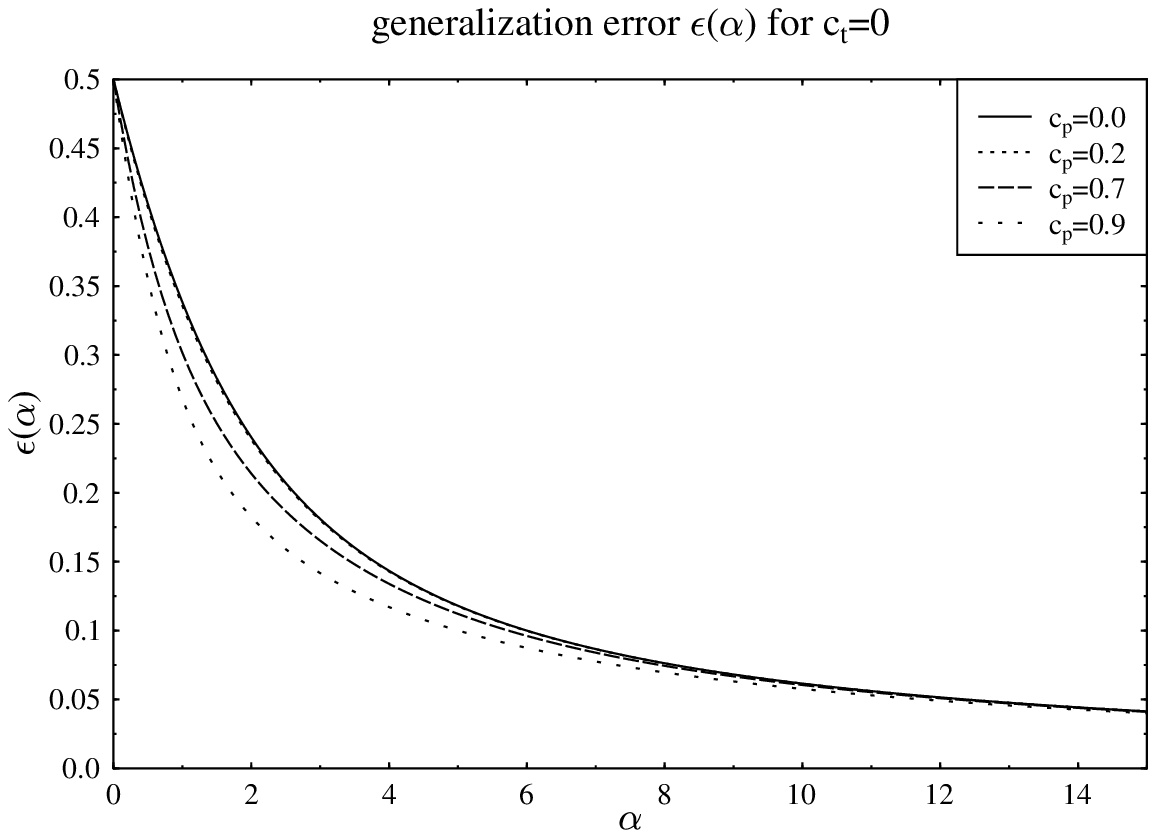}
\centerline{\underbar{Fig.~3}}
\epsfxsize=15cm
\epsfbox{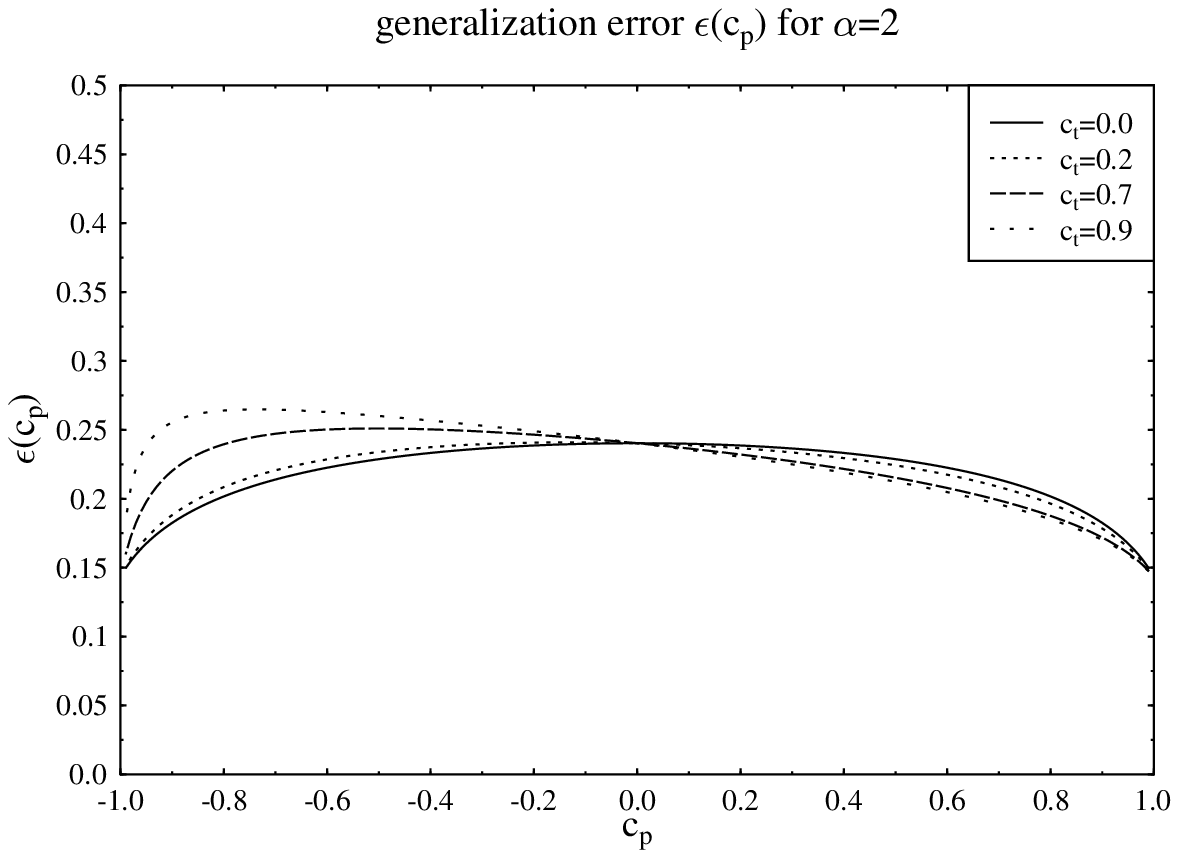} 
 \centerline{\underbar{Fig.~4}} 
\epsfxsize=15cm
\epsfbox{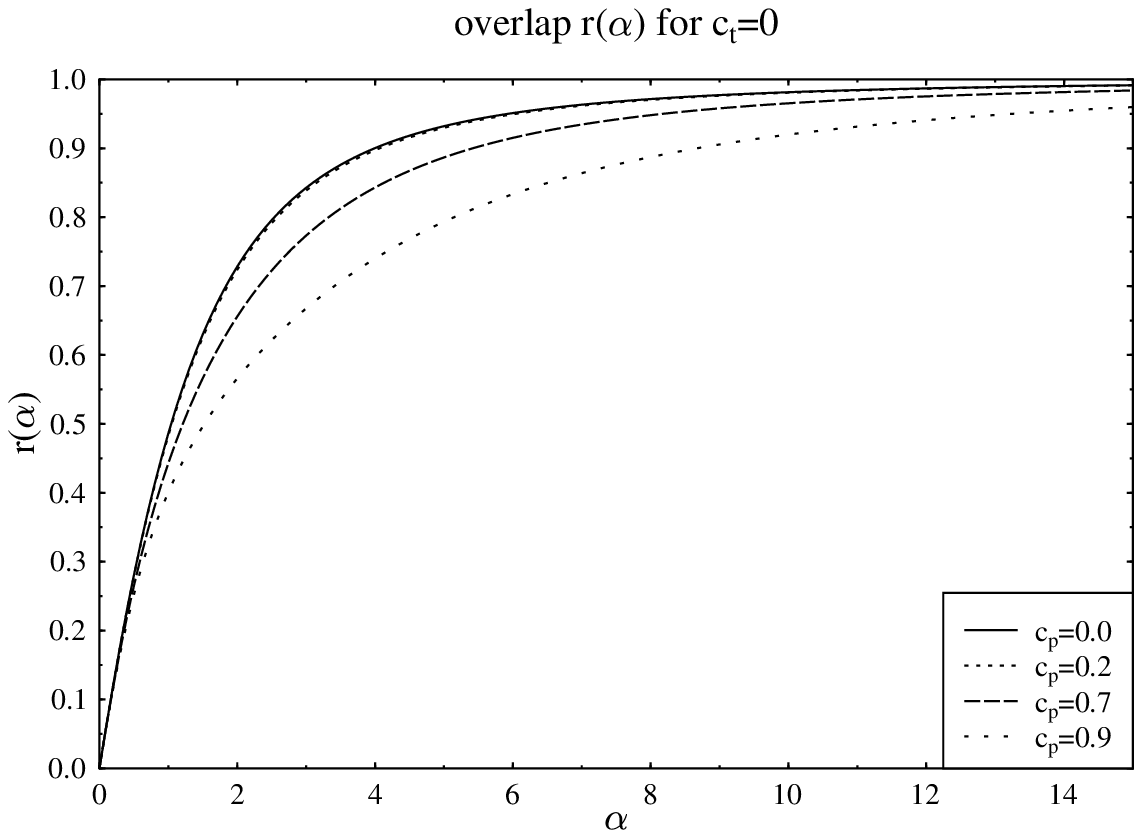}
 \centerline{\underbar{Fig.~5a}}
 \epsfxsize=15cm
\epsfbox{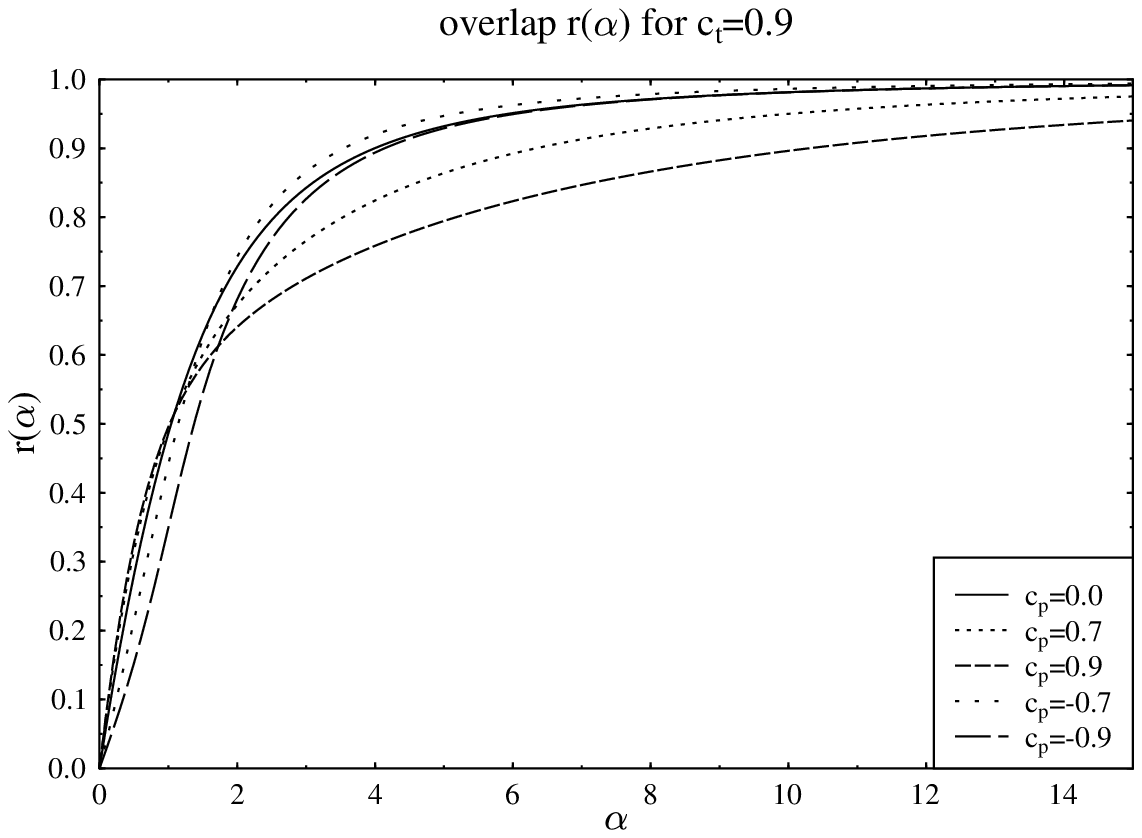}
 \centerline{\underbar{Fig.~5b}}
\epsfxsize=15cm
\epsfbox{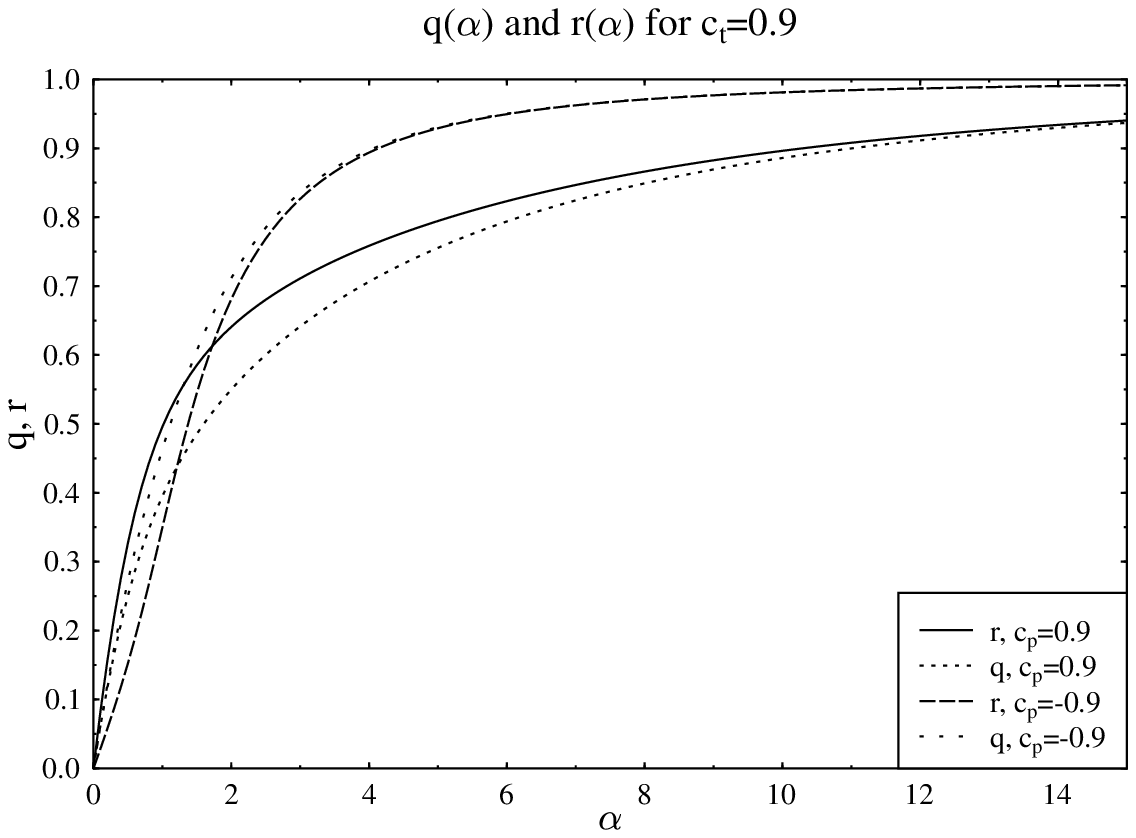}
 \centerline{\underbar{Fig.~6}}
 \epsfxsize=15cm
\epsfbox{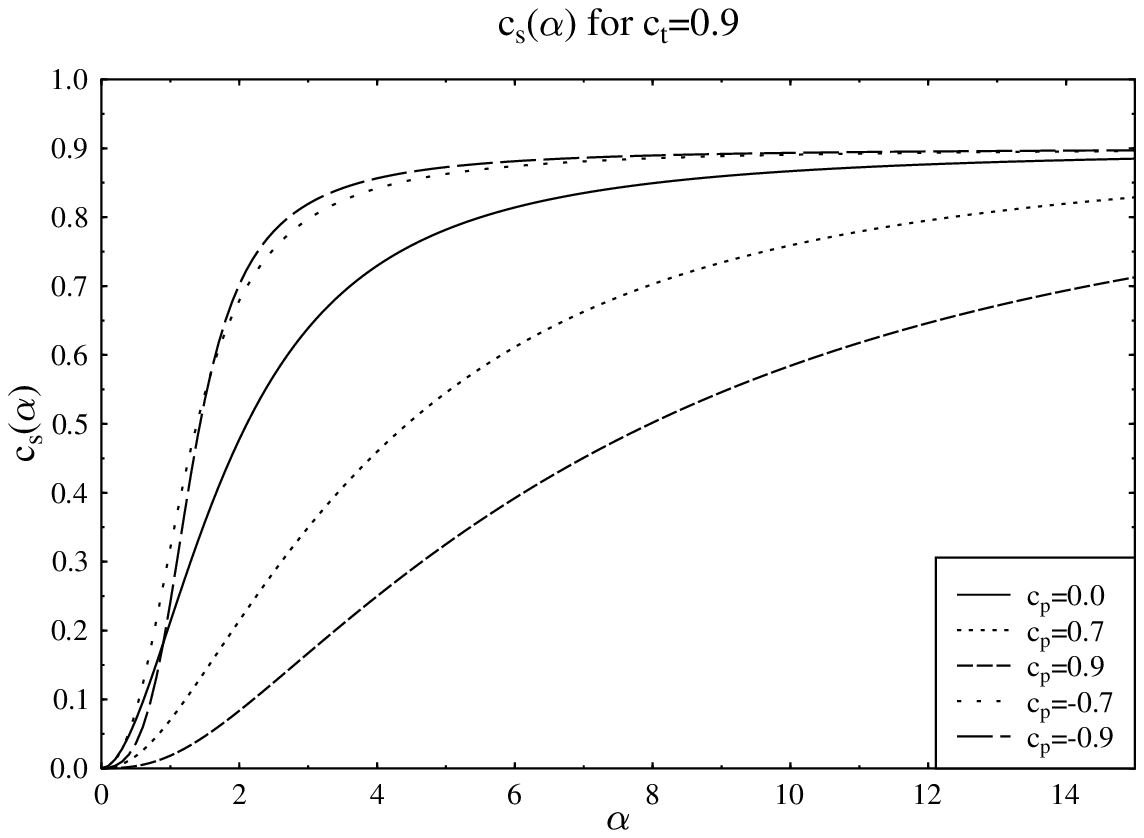}
 \centerline{\underbar{Fig.~7}}
\epsfxsize=15cm
\epsfbox{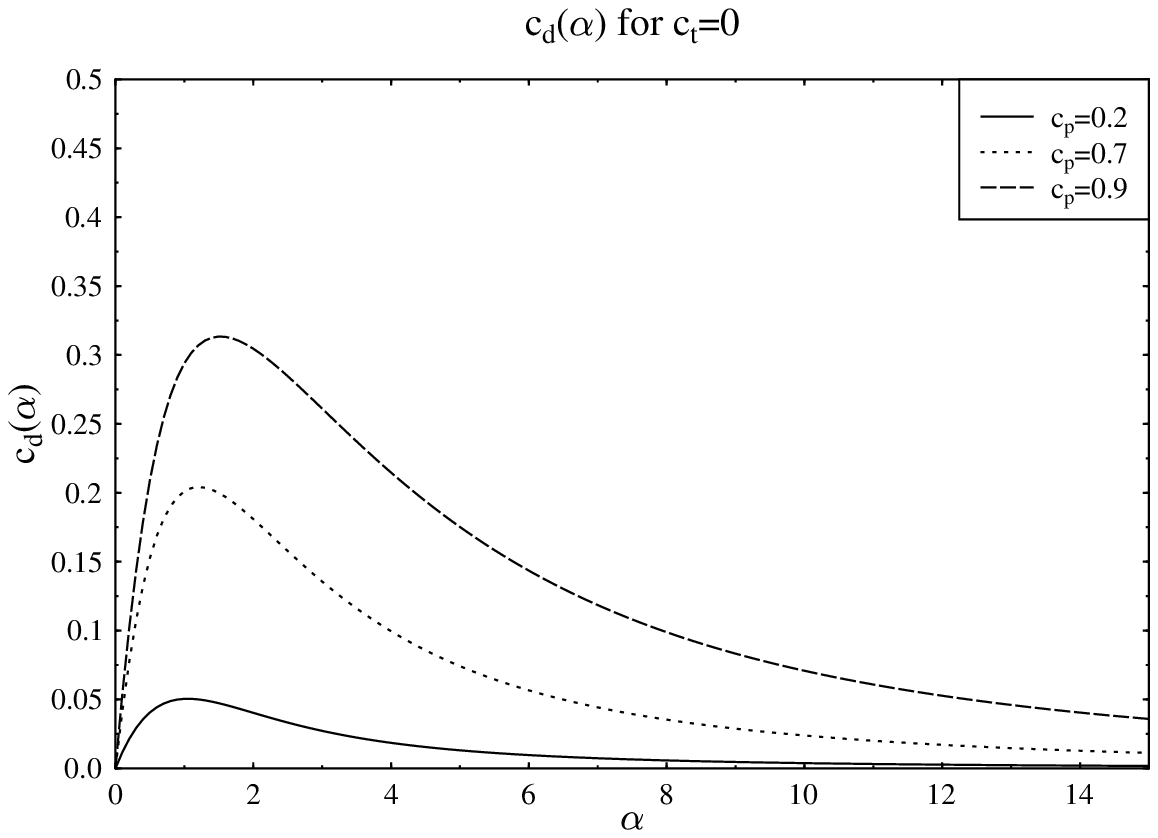}
 \centerline{\underbar{Fig.~8}}
\epsfxsize=15cm
\epsfbox{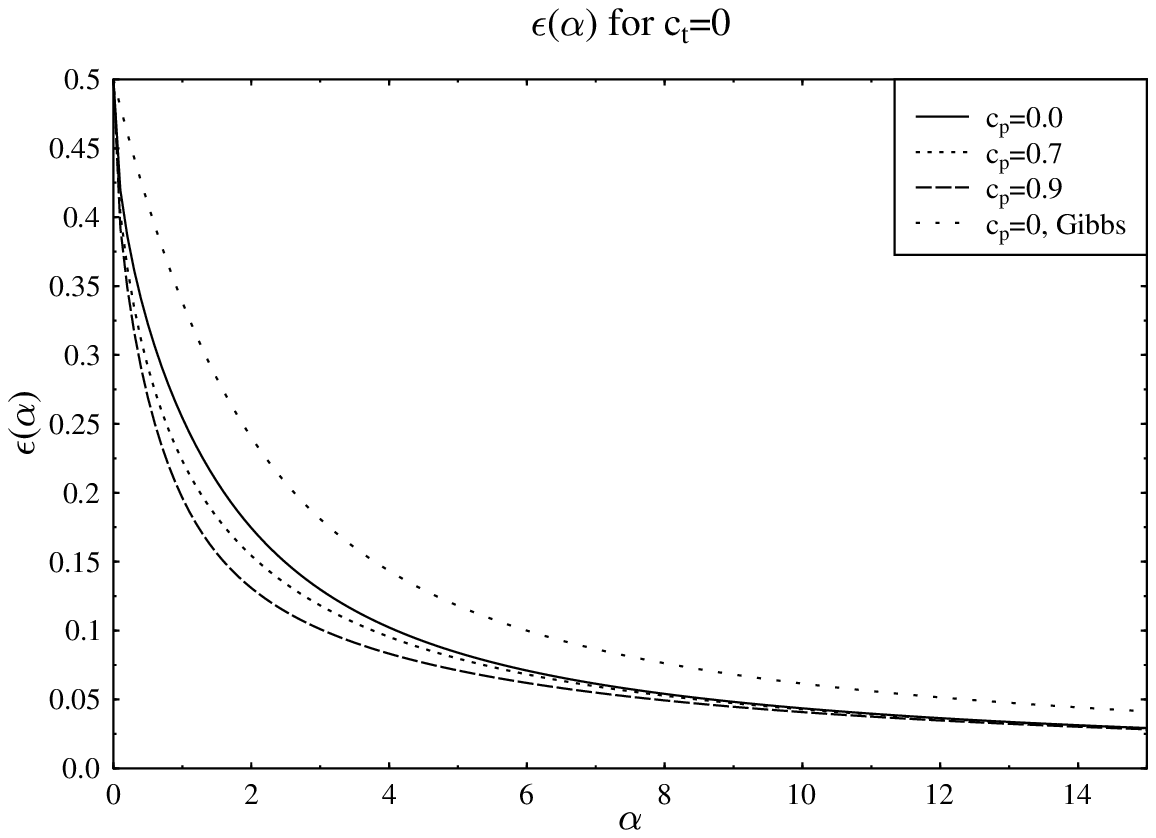}
 \centerline{\underbar{Fig.~9}}
\epsfxsize=15cm
\epsfbox{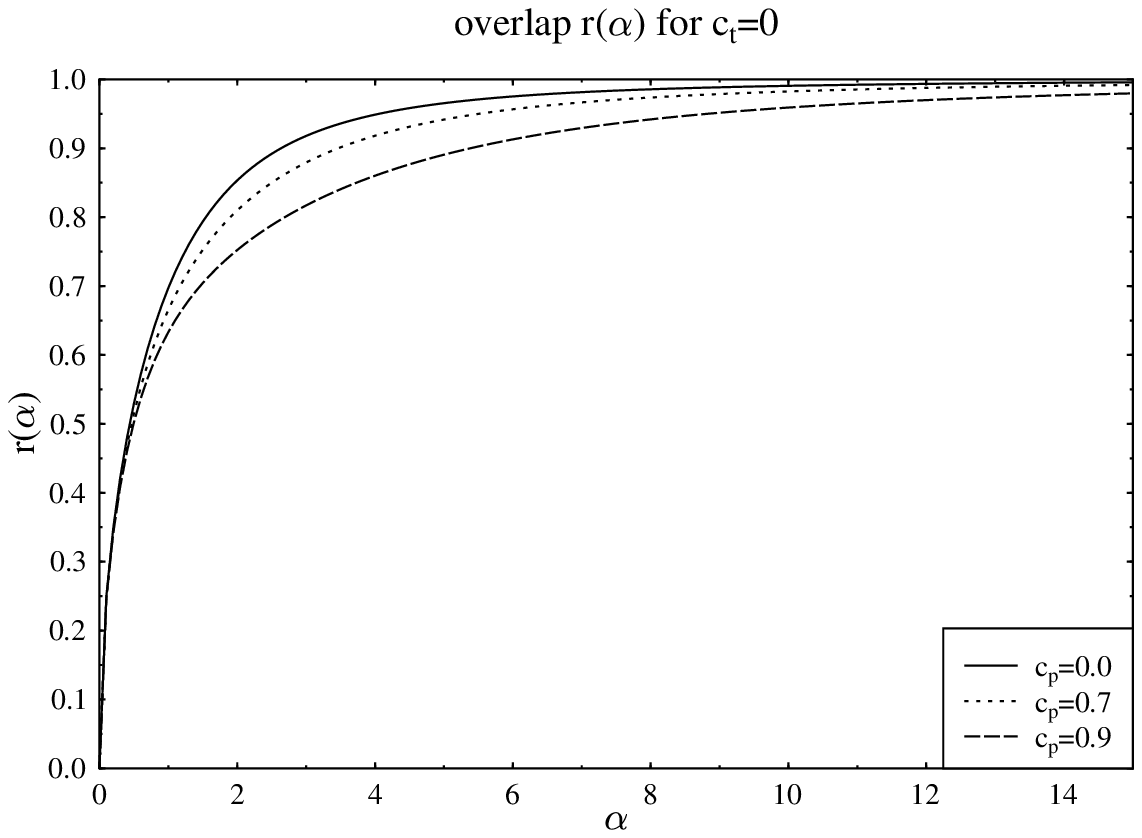}
 \centerline{\underbar{Fig.~10a}}
\epsfxsize=15cm
\epsfbox{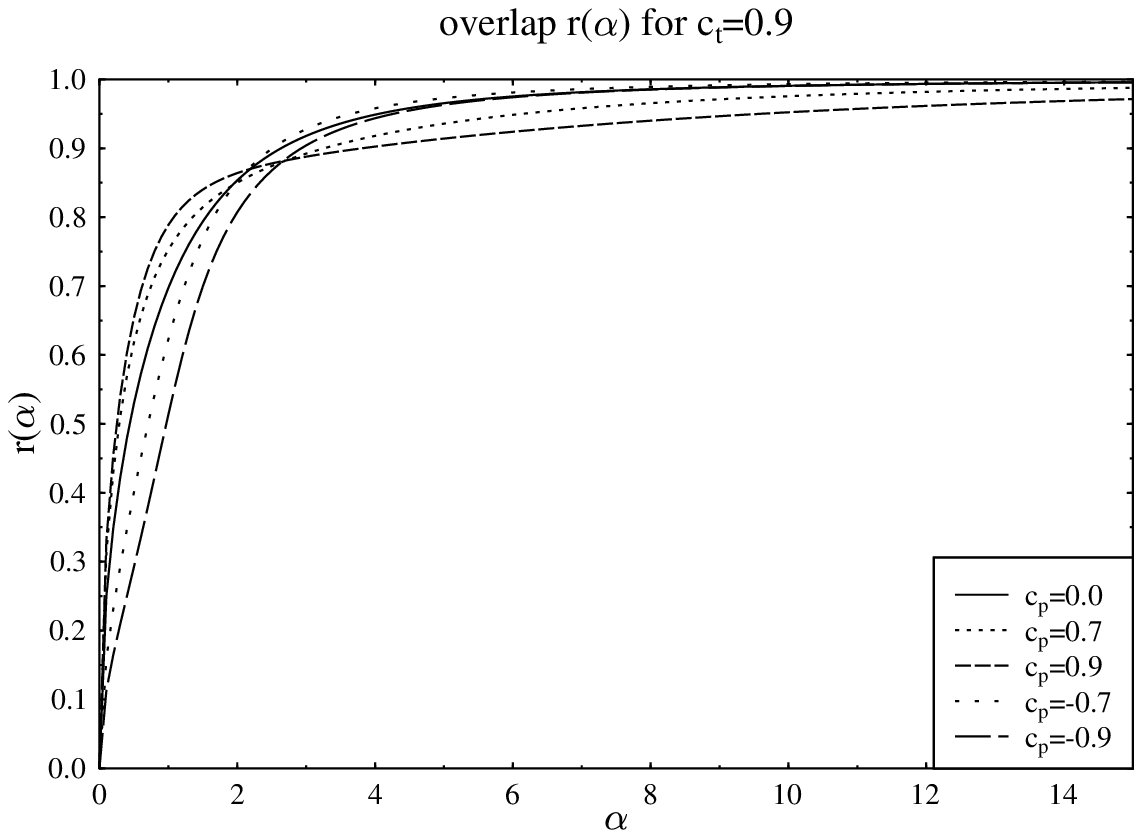} 
 \centerline{\underbar{Fig.~10b}}
 \epsfxsize=15cm
\epsfbox{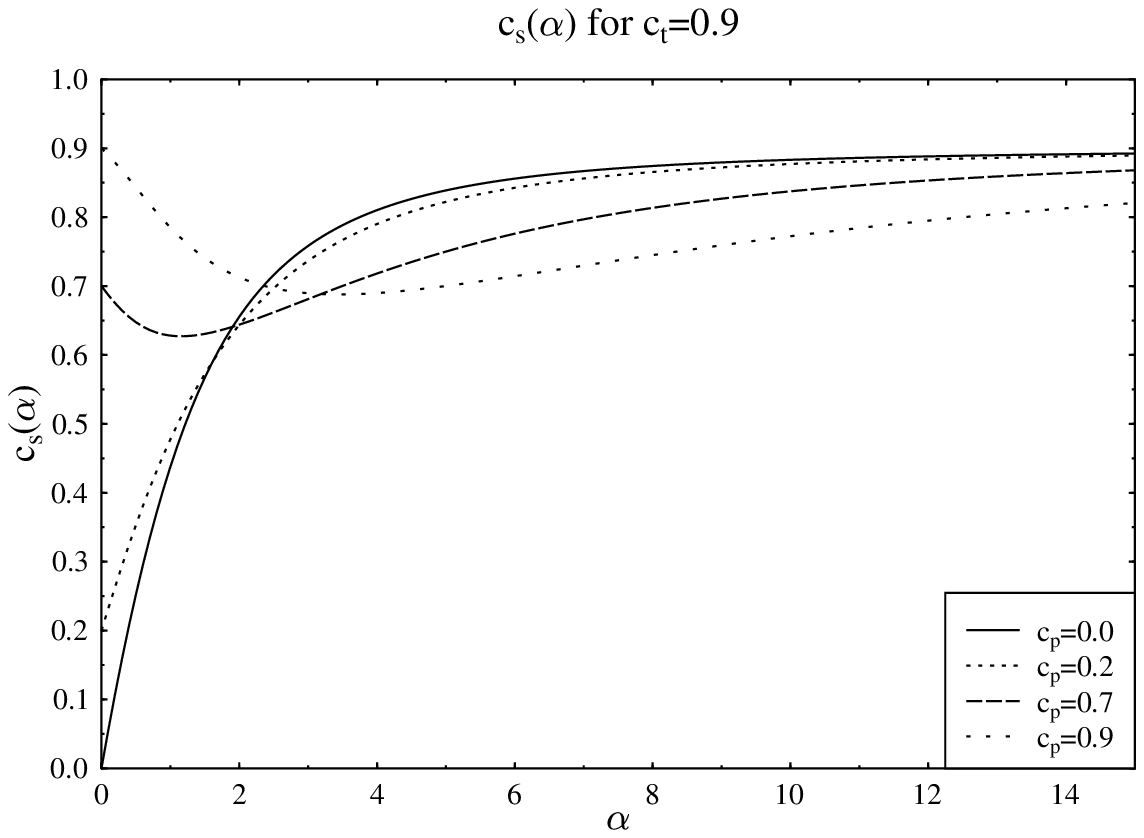}
 \centerline{\underbar{Fig.~11}}
 \epsfxsize=15cm
\epsfbox{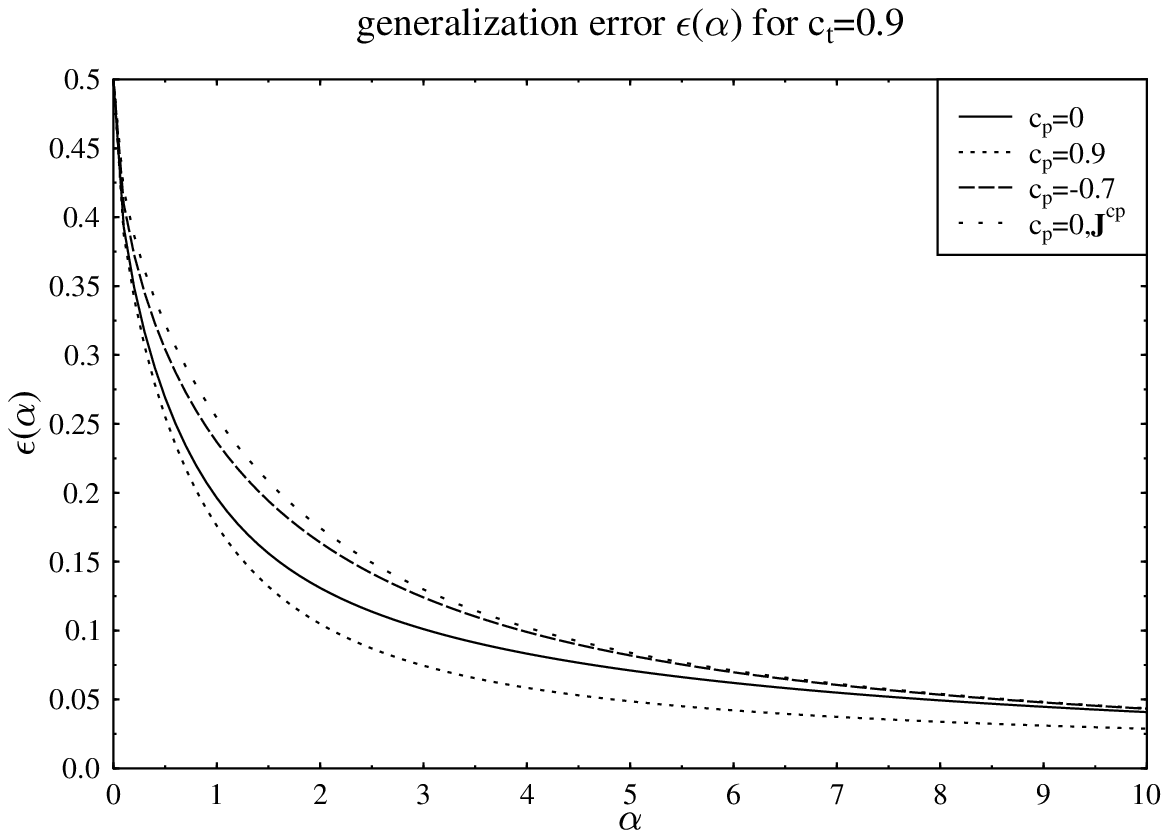}
 \centerline{\underbar{Fig.~12}}
}
\end{document}